# Structure of poly(propyl ether imine ) (PETIM) dendrimer from fully atomistic molecular Dynamics Simulation and by Small Angle X-ray scattering

Chandan Jana[1], G. Jayamurugan[1], Rajesh Ganapathy[3], Prabal K. Maiti[2*], N. Jayaraman[1*] and A. K. Sood[3*]

[1]Department Organic Chemistry, Indian Institute of Science, Bangalore, India 560012

[2]Center for Condensed Matter Theory, Department of Physics, Indian Institute of Science, Bangalore, India 560012

[3]Department of Physics, Indian Institute of Science, Bangalore, India 560012

*Abstract:*

We study the structure of carboxylic acid terminated neutral poly (propyl ether imine) (PETIM) dendrimer from generation 1 through 6 (G1-G6) in a good solvent (water) by fully atomistic molecular dynamics (MD) simulations. We determine as a function of generation such structural properties as: radius of gyration, shape tensor, asphericity, fractal dimension, monomer density distribution, and end-group distribution functions. The sizes obtained from the MD simulations have been validated by Small Angle X-Ray Scattering (SAXS) experiment on dendrimer of generation 2 to 4 (G2 –G4). A good agreement between the experimental and theoretical value of radius of gyration has been observed. We find a linear increase in radius of gyration with the generation. In contrast, $R_g$ scales as $\sim N^x$ with the number of monomers. We find two distinct exponents depending on the generations: $x = 0.47$ for G1-G3 and $x = 0.28$ for G3-G6 which reveals their non-space filling nature. In comparison with the amine terminated PAMAM dendrimer, we find $R_g$ of G-th generation PETIM dendrimer is nearly equal to that of (G+1)-th generation of PAMAM dendrimer as observed by Maiti et. al. [Macromolecules, **38**, 979 2005]. We find substantial back folding of the outer sub generations into the interior of the dendrimer. Due to their highly flexible nature of the repeating branch units, the shape of the PETIM dendrimer deviates significantly from the spherical shape and the molecules become more and more spherical as the generation increases. The interior of the dendrimer is quite open with internal cavities available for accommodating guest molecules suggesting using PETIM

---
[*]For correspondence:  maiti@physics.iisc.ernet.in  (PKM) or  jayaraman@orgchem.iisc.ernet.in (NJ)or asood@physics.iisc.ernet.in (AKS)

dendrimer for guest-host applications. We also give a quantitative measure of the number of water molecules present inside the dendrimer.

**I.** *Introduction:*

Dendrimer,[1] regularly branched polymers, of different initiator core, branches and the peripheral groups have been synthesized [2-8] in the recent decades. Numerous kinds of experiment have been carried out to prove the potential application of this type of new material in biochemical, medical, technical and industrial fields[9-12]. In last two decades lots of efforts have been made to design and synthesize biocompatible dendrimer with different monomers having variety of functionalities. The atomistic characterization of dendrimer structure has lagged this rapid progress in synthesis and design [13]. The problem is that these molecules possess an enormous number of energetically permissible conformations, and in solution there is rapid interchange between them. Thus diffraction techniques yield little structure information. Also many generations involve the same monomers, making it difficult to extract precise information about the local structure. Thus the first precise experimental data about the gross size came from size exclusion chromatography (SEC), which is now being complemented with Small Angle Neutron Scattering (SANS) and Small Angle X-ray Scattering (SAXS) to determine gross size and some structural details of dendrimers. [14-16] In recent years, computational and theoretical techniques [17-35] proved to be very effective elucidating the structural properties of the dendrimer both in good and bad solvent[18, 19]. Many of these theoretical results agree very well with the available experimental data[36-37, 15, 16] on such systems.

Recently Jayaraman et. al. [38] have synthesized PETIM dendrimer and reported their cytotoxic properties. Evaluation of the extent of cytotoxicity indicates that the toxicity levels of these dendrimer are very mild and this point to the possibility of using PETIM dendrimer in various applications. However, so far no structural information is available for these dendrimer, which is essential for their application ranging from drug delivery to molecular encapsulations. Here we report a comprehensive understanding of the structure of PETIM dendrimer using fully atomistic simulations and have used SAXS to validate some our simulation results.

Following is the outline of the paper. In section 2.1 we describe the structure and the atomistic feature of the PETIM dendrimer, in section 2.2 the experimental details of the SAXS, and in section 2.3 the details of building of the atomistic model and simulation methods. In section 3 we have summarized the results obtained from the simulation as well as SAXS studies. Finally, the summary of the main results and the conclusions are given in section 4.

## 2. *System and Methods:*

### *2.1 Samples:*

The synthesis and cytotoxicity studies of carboxylic acid terminated Poly(Propyl Ether Imine) (PETIM) dendrimer (up to generation 3) reported[38] recently proves its excellent biocompatibility. PETIM dendrimer starts growing (**Figure 1**) three dimensionally from the oxygen as the core and branches out at each tertiary nitrogen, which is separated by eight-bond spacer for each generation of the dendrimer. The spacer containing all the $SP^3$ hybridized atoms of the PETIM dendrimer is flexible enough to have numerous numbers of molecular conformations. As the spacer length of the PETIM dendrimer (8 bonds) is larger than PAMAM (7 bonds) and PPI (4 bonds) dendrimer, the size of PETIM are expected to be larger than PAMAM dendrimer as well as PPI dendrimer for the same number of terminal groups and generations. Due to its larger spacer length, the compactness and space filling nature of the molecule is expected to be different compared to PAMAM and PPI dendrimers where the branching occurs at shorter distances. The fully atomistic description of the PETIM dendrimer has been described in *Table 1*.

**2.2 Small angle X-Ray scattering:**

**Experimental Section:**

The samples were loaded in thin quartz capillary and were subjected to Cu K$\alpha$ (0.154 nm) radiation from a sealed tube x-ray generator (Philips, PW3830) with a line focus beam. A small angle resolution of 0.01 Å$^{-1}$ was achieved using a small angle Kratky camera (Hecus M. Braun, SWAXS) with line collimation. A slit size of 200μm and a beam width of 10mm were used in all our experiments. The instrument resolution (FWHM) was found to be 0.08 nm$^{-1}$. A 1-D position sensitive detector (MBraun PS D50M) was used to measure the scattered intensity. An adjustable sample holder with temperature regulation unit (Anton PAAR, KHR) was used to maintain the temperature of the samples at 28$^{o}$C. The sample chamber was evacuated to ~ 0.01 mbar to prevent stray scattering. Typical exposure times were about 8 hrs.

**Theoretical Background:**

SAXS is a well-established technique for obtaining information about the size, shape and structural correlations of macromolecules in solution. The scattered intensity from a collection of macromolecules is given by the expression

$$I(q) = A P(q) S(q) \qquad (1)$$

where q is the wave vector transfer and is given by $q = 4\pi \sin\theta/\lambda$, $\theta$ is the scattering angle, $\lambda$ is the wavelength of the radiation used, A is the scattering amplitude which depends on the number density and electron density contrast, P(q) is the particle form factor and has information regarding the size and shape of the macromolecule and *S(q)* is the structure factor having information regarding inter-particle correlations. In a dilute collection of macromolecules *S(q)* ~ *1* and equation (1) reduces to

$$I(q) = AP(q) \qquad (2)$$

For a collection of dilute spheres with radius of gyration $R_g$ the scattered intensity *I(q)* in the low q limit $q R_g << 1$ is given by the Guinier relation

$$I(q) = I(0)\exp\left[-\frac{q^2 R_g^2}{3}\right] \qquad (3)$$

A log-linear plot of *I(q)* vs. $q^2$ in the low q limit will result in a straight line and its slope yields the radius of gyration $R_g$.

*2.3 Simulation:*

The initial three-dimensional molecular models of various generations PETIM dendrimer were built as follows: First the core fragment, the branching repeat unit and terminal repeat unit was built and energy minimized using 3D-sketure of the Cerius2 software.[39] The partial charges on the atoms on each fragment were derived using the charge equilibration (QEq) method[40]. Then the fragments were assembled properly again using 3D-sketure of the Cerius2 software to make a given generation and was optimized for 5000 steps using conjugate gradient minimization. Dreiding force field[41] was used to describe the inter-atomic interactions. The charge equilibrated minimized structures of all generations have been directed through the annealing algorithm using OFF module of Cerius2 software as follows:

The initial minimized structure was heated at a rate of 100K/4 ps from 300K to 2000K, followed by quenching to 1000K at the same rate, followed by 4 such cycles between 1000 and 2000K, and finally cooling to 300K. The annealed structures were solvated with TIP3P water model using xleap module of AMBER7[42]. The box dimensions were chosen in order to ensure a 10Å solvation shell around the dendrimer structure. This procedure resulted in solvated structures, containing between 3301 atoms for G1 and 67669 atoms for G6. MD simulation was performed using the AMBER7 [42] software suite, using the Dreiding force field.[41] The solvated structures were subjected to 1000 steps of steepest descent minimization of potential energy, followed by another 2000 steps of conjugate gradient minimization. During this minimization the dendrimer structure was kept fixed in their starting conformations using a harmonic constraint with a force constant of 500 kcal/mol/Å$^2$. This allowed the reorganization of the water molecules to eliminate bad contacts with the dendrimer structure. The minimized structure was then subjected to 45 ps of MD, with 2 fs time step. During the dynamics, the system was gradually heated from 0 to 300 K with harmonic constraints on the solute using the SHAKE method. This was followed by 200 ps constant volume – constant temperature (NVT) dynamics with a temperature-coupling constant of 0.5-1.0 ps on the solute. Finally, 2-10 ns (depending on the generations) NPT unrestrained production dynamics was carried out with a time constant for heat bath coupling of 1ps. The electrostatics interactions were evaluated with the Particle Mesh Ewald [43] (PME) method, using a real space cut off of 9Å. This simulation times proved to be long enough to get equilibrium properties as seen from the time evolution the radius of gyration of the dendrimer shown in Figure 2.

## 3. Results and Discussion:

To characterize the structure and properties of dendrimers as a function generation we have chosen the following quantities: radius of gyration, shape tensor, asphericity, monomer density distribution, molecular surface area, end group distribution, solvent accessible surface, molecular volume, and spatial arrangement of branch points. We have also done SAXS experiment to compare our simulation results with those obtained from experiment. We have also studied the penetration of water inside dendrimer by computing the number of waters inside the dendrimer as a function of the distance form the center-of-mass of the dendrimer.

**Size and Shape:**

To obtain a quantitative estimate of the average size of the dendrimer, mean square radius of gyration $<R_g^2>$ defined as

$$\langle R^2_g \rangle = (1/M) \left\langle \left[ \sum_{i=1}^{N} m_i \, |r_i - R|^2 \right] \right\rangle \quad (4)$$

has been computed ($R_g(a)$ given in Table 2) over the trajectory. Here $R$ is the center-of-mass of the dendrimer, $m_i$, $r_i$ are the mass and position vector of the $i$-th atom, and $M$, $N$ are the total mass and total number of atoms of the dendrimer. Table 2 shows the radius of gyration (denoted as $R_g(a)$) obtained from our simulation as a function of generation. To compare our simulation results we have also shown the size obtained from our SAXS studies, details of which is given the next section.

The radius of gyration of the dendrimers in solution has been determined from the Small angle X-ray scattering (SAXS). Fig 3 shows the Guinier plots ($I(q)$ vs. $q^2$) for three different generations of the dendrimers after subtracting the background due to the solvent and the capillary. The black lines are the fits to the data using eq. (3). The concentration of the third generation (Fig 3b) and fourth generation (Fig 3c) dendrimers was 0.5wt% and for the second generation (Fig 3a) was 1.5wt% to have a good signal to noise ratio. In spite of this three fold increase in concentration over the third and fourth generation dendrimers, the scattering from the second generation was low and hence the data shown for the second generation was smoothened using a 5-point adjacent-averaging method. The values of $R_g$ obtained from Guinier plot for generations G2 – G4 have been tabulated in Table 2. Closely related to the results of SAXS experiments is the spherically averaged Fourier Transform of the single particle density, $I(q)$, given by Equation 5 [22]

$$I(q) = \left(1/4\pi N^2\right) \int_0^{2\pi} d\phi \int_0^{\pi} \sin\theta \, d\theta \, |\sum_{i=1}^{N} \exp\left[i\vec{q}\cdot\vec{r_i}\right]|^2 \qquad (5)$$

Where

$$\vec{q} = q\sin\theta\cos\phi\,\hat{x} + q\sin\theta\sin\phi\,\hat{y} + q\cos\theta\,\hat{z}$$

is the scattering vector and $r_i$ is the position vector of the scattering center.

To make comparison with the SAXS data, we have calculated $I(q)$ using orientation averaging at intervals of 9º in both θ (0< θ < 180) and φ (0< φ < 360). Figure 4 shows the plot of ln(I(q)) vs $q^2$ as obtained from our simulation data. Linear regression fit to the theoretical Guinier plots obtained this way (ln(I(q)) vs $q^2$ ) gives radius of gyration $R_g$ for each generation (denoted by $R_g(s)$ in Table 2) and has been shown in Figure 5 as a function of generation. Table 2 shows variation of $R_g$ as a function of generation obtained from three different ways and we find a good agreement between our calculated $R_g$ with the values obtained the SAXS data.

Figure 5 show that the radius of gyration increases linearly having slightly different slopes with the generation number for all the three cases. This linear dependence of the radius of gyration on the generation has been observed for PPI dendrimer[17]. We also see that our calculated values for $R_g$ are consistently higher than that of experimental one. This discrepancy could be attributed to the fact that the shape of the dendrimer (discussed below) is far from spherical, an assumption that was made to extract $R_g$ when using Eq. 3. The discrepancy between the experimental and simulation values may also be due to the solution conditions. The experiments were conducted in methanol solution, whereas the simulations were performed in water. The consistent higher $R_g$ values for the aqueous solutions indicate that the propylene spacers in dendrimers are stretched under this solution condition. In methanol solutions, the propylene spacers experience relatively less hydrophilic environments and thus less affected by the solution condition. The lesser $R_g$ values of dendrimers in methanol solutions reflect the extent of shrinking of the dendrimers under this solution condition. Swelling in aqueous solutions and shrinking in methanol solutions may contribute in addition to the asphericity for unequal $R_g$ values in these two solvents.

The variation of root mean square radius of gyration with number of monomers shown in Figure 6 follows the scaling relation $\langle R_g \rangle \sim N^\alpha$ with $\alpha = 0.47$ for G1-G3 and $\alpha = 0.28$ for G3-G6. It is clear that a single law $\langle R_g \rangle \sim N^\alpha$ does not describe the $R_g$ dependence in the whole range of $N$ studied here and far from the universal power law of the type $\langle R_g \rangle \sim N^{0.33}$ previously obtained for PAMAM and PPI dendrimer both theoretically and experimentally[18, 19, 44]. Such non-universal scaling law behavior was recently found to be true for flexible dendrimers[45]. As the generation number increases the dendrimer structure becomes more and more compact and space filling and the exponent is approaching the limiting value of 0.33. The above scaling exponent can be used to calculate the fractal dimension ($d_f = 1/\alpha$) of the dendrimer from the relation

$$N \propto R_g^{1/\alpha} \qquad (6)$$

We find the fractal dimension $d_f$ to be 2.1 for G1-G3 and 3.4 for G3-G6 respectively. PAMAM dendrimer shows this exponent as 3.0[19], which is equal to the dimensionality of the space indicating their space filling and compact structures. For lower generation PETIM dendrimer (G1-G3), the fractal dimension (~2.1) is far from the dimensionality of the space, which indicates that these molecules are non-space filling and open in nature. For higher generation of the PETIM dendrimer, fractal dimension goes near to the dimensionality of the space; still the spatial arrangement of the branches within the molecule remain non-space filling. This non-space filling and open nature of the PETIM dendrimer have been further confirmed from the calculation of the single particle Form factor (see Kratky plots shown in Figure S1 in the supplementary materials). Due to larger and flexible nature of the spacer the molecule gets larger span of space to orient themselves resulting in their non –space filling nature.

Finally to make a comparison with the available data on other types of dendrimer we find that the radius of gyration of PETIM dendrimer is larger than that of PAMAM dendrimer for a given generation. To a good approximation radius gyration of G-th generation of PETIM dendrimer is approximately greater or equal to that of G+1-th generation (observed by Maiti et. al.)[19] and G+2-th generation of the PAMAM dendrimer (observed by M. Han et. al.)[35]. Again this is a consequence of larger and more flexible nature of the spacer in PTEIM dendrimer compared to other dendrimers like PAMAM and PPI.

**Shape:**

The shape of PAMAM dendrimer has been studied extensively using transmission electron microscopy (TEM)[46] and with tapping mode atomic force microscopy (AFM).[47] However, so far there is no data on the shape of the PTEIM dendrimer. The flexibility and larger length of the repeat unit determine the shape of these molecules. To visualize the variation of the shape as a function of generation we have shown a snap shot of the final configuration for each generation G1 to G6 in Figure 7. To provide a more quantitative criteria the aspect ratio, which is the ratio of the two principle moments of inertia, have been calculated and averaged over the dynamics trajectory. $I_x$, $I_y$ and $I_z$ are three principle moments of inertia of the molecule, which have been found by diagonalizing the gyration metrics G.[48]

$$G_{mn} = \frac{1}{M}\left[\sum_i^N m_i (r_{mi} - R_m)(r_{ni} - R_n)\right] \qquad m,n = x,y,z \qquad (7)$$

where $r$ and $R$ are the coordinates of atoms and the center of mass of the dendrimer respectively. $m_i$ is the mass of ith atom. $M$ is the total mass of the dendrimer. The average values of the three principal moments of inertia are tabulated in Table 4, while Figure 8 (a) shows the average ratios for different generations dendrimer. We see that $I_z/I_x$ and $I_z/I_y$ vary from 20.163-2.372 and 5.54-1.58 respectively from generation 1 to 6 (Table 4).

As the generation increases the value of the aspect ratio goes towards the 1.0 (Table 4) indicating that the shape of the dendrimer approaches to the spherical one. More quantitative information about the shape of the dendrimer can be obtained by calculating the *asphericity,* introduced by Rudinck and Gaspari as [49]

$$\delta = 1 - 3\left(\frac{\langle I_2 \rangle}{\langle I_1^2 \rangle}\right) \qquad (8)$$

Where $I_1$, $I_2$ and $I_3$ are defined as

$$I_1 = I_x + I_y + I_z$$
$$I_2 = I_x I_y + I_y I_z + I_x I_z$$
$$I_3 = I_x I_y I_z$$

The relative shape anisotropy of the simulated dendrimer shown in Figure 8 (b) shows that with the increase in generation the dendrimer becomes more and more compact spherical structure. The asphericity decreases from 0.4 for G1 to 0.05 for G6.

**Radial monomer density profiles**

The average radial monomer density ρ(r) can be defined by counting the number $N(R)$ of atoms whose centers of mass are located within the spherical shell of radius r and thickness Δr. Hence, the integration over r yields the total number of atoms as:

$$N(R) = 4\pi \int_0^R r^2 \rho(r) dr$$

In Figure 9 we show the radial monomer density for various generation PTEIM dendrimers in water. In each case the plot shows the contributions to a particular generation from each of its component generations. We take the origin as the center of mass. We see a very high-density region around the origin and a tailing zone in which the monomer density is gradually decreasing with the radial distance. This indicates that the core region is very dense compared to the middle of the dendrimer, which is fairly hollow supporting the dense-core picture from earlier theoretical and computational studies.[13, 22, 31] However, there is no constant density region in monomer density distribution in the middle of the dendrimer as has been observed for PAMAM dendrimers.[18] The radial monomer density distribution for each sub-generation shows how the inner sub-generations are distributed throughout the interior of the dendrimer and indicates there is significant back folding of the outer sub generation. The extent of back folding increases with the increase in generations. It is clear that the monomer density is higher at the core region for all the generation compared to the exterior of the molecule. On the basis of this observation it can be concluded that the sub-generation for a particular generation has folded back towards the core and makes the core region compact compared to the region far from the core. With increase in the generation number the monomer density increases and extends radially outward from the core. This dense core picture is in agreement with the results obtained by Boris and Rubinstein[21] for the dendrimer containing the flexible repeat unit.

**Water Penetration:**

In a good solvent like water PTEIM has lots of internal voids and cavities, which can act as a binding site for small molecules for drug delivery and skin care products and these cavities can accommodate a large number of water molecules as well. Due to the favorable interaction of water with the various functional groups of the dendrimer significant number of water penetrates inside the dendrimer and it helps swelling the dendrimer. A quantitative estimate of the solvent penetration is given by counting the number of waters bound by the dendrimer outer surface. Due to the non-uniformity as well as asphericity of the dendrimer surface special care must be taken to identify the bound water, as simple spherical cutoff will overestimate the numbers of waters within the dendrimer. To have an accurate estimate of the number of bound water we have used following criteria[19, 50]: We first calculated the molecular surface area (MSA) for each of the dendrimer atom using a large probe radius (6 Å). With this probe radius the generated surface of the dendrimer becomes almost spherical and smooth. Those atoms with non-zero MSA represent the surface atoms of the dendrimer. Using these surface atoms we identify all the surface waters that are within 4 Å of the surface atoms. Next we identify all the waters close to the inner atoms (with zero MSA) excluding all the previously defined surface waters. The number of bound waters calculated this way is listed in Table 5. This significant penetration of solvent molecules inside the dendrimer structure is in agreement with the recent SANS studies on poly (benzyl ether) [51] and polycarbosilane dendrimers [52]. In these experiments the number of solvent molecules inside the dendrimer was calculated from the change in neutron scattering density. We find that the number of bound water for the G-th generation of the PETIM is larger than the number of water in same generation PAMAM dendrimer. The number difference of bound water between Gth generation of PETIM and G+1th generation of PAMAM increases as we go to higher generation. For example the number difference between PETIM-32 and PAMAM-64 is about 14%, where as for PETIM-128 and PAMAM-256 is 30%. So for various applications lower generation PETIM dendrimer can be used in place of higher generation PAMAM to avoid some extra toxicity towards the living cell.

**4.** *Conclusion:*

SAXS has been used to study the size of the carboxylic acid terminated neutral poly (propyl ether imine) (PETIM) dendrimer for generation two to four (G2-G4) in water. The size obtained from the SAXS measurement is in good agreement with the atomistic molecular dynamics simulation in explicit water. The dependence of the radius of gyration as a function of the number of monomers $N$ does not obey the scaling law $R_g \sim N^{1/3}$ as has been observed for PAMAM and PPI dendrimer. Instead for higher generation PTEIM dendrimer studied here (up to G6) we find the scaling form to be $R_g \sim N^{0.28}$. The monomer density distribution shows the dense core nature of the dendrimer, which is expected for dendrimer having flexible repeat unit. Significant back folding is observed for all the generations studied. This along with our previous simulation studies on PAMAM dendrimer demonstrates that back folding is universal phenomena for dendrimer architecture. With the very flexible repeat unit the shape of the dendrimer is far from spherical one. We find significant penetration of solvent molecules in the interior of dendrimer molecules for all generations.

**Supporting information Available:**

1) The xyz coordinates for a snapshot of the trajectory for each dendrimer from Generation 1 to Generation 6 in asci format.
2) The monomer density profiles with respect to the center of core of the dendrimer.
3) Branch point distribution
4) Solvent accessible surface area and volume

**Acknowledgement** NJ thanks Department of Science and Technology, New Delhi, for a financial support. GJ thanks Council of Scientific and Industrial Research, New Delhi, for a research fellowship. We thank SERC, IISc, Bangalore, for generous computer time where all the computations have been carried out.

*Table1:* *Atomistic description of PETIM dendrimer.*

| Generations (G) | No of terminal group ($N_t$) | Molecular Weight (g/mol) | Total No of Atoms ($N$) | Total No of Nitrogen Atom ($N_n$) |
|---|---|---|---|---|
| 1 | 4 | 420.4547 | 61 | 2 |
| 2 | 8 | 1169.4149 | 181 | 6 |
| 3 | 16 | 2667.3269 | 421 | 14 |
| 4 | 32 | 5663.1509 | 901 | 30 |
| 5 | 64 | 11654.7988 | 1861 | 62 |
| 6 | 128 | 23638.0938 | 3781 | 126 |

*Table2:* *Table shows the radius of gyration ($R_g$) obtained from three different way as the function of generation of PETIM dendrimer. $R_g(a)$ = radius of gyration calculated from the equation 4. $R_g(s)$ = radius of gyration obtained from the theoretical Guinier plot. $R_g(SAXS)$ = radius of gyration obtained from the experimental Guinier plot. $R_T$ = radius of gyration calculated considering the terminal groups only. The quantities have been averaged over 500 ps after the equilibration.*

| No terminal groups | $R_g(a)(Å)$ | $R_T(PETIM)(Å)$ | $R_g(s)(Å)$ | $R_g(SAXS)(Å)$ |
|---|---|---|---|---|
| 4 | 5.65 ± 0.27 | 6.59 | 5.19 ± 0.12 | |
| 8 | 8.90 ± 0.37 | 10.64 | 8.19 ± 0.36 | 7.86 ± 0.1 |
| 16 | 14.14 ± 0.39 | 16.88 | 13.18 ± 0.24 | 10.74 ± 0.16 |
| 32 | 17.73 ± 0.41 | 20.58 | 16.10 ± 0.53 | 14.88 ± 0.21 |
| 64 | 21.12 ± 0.60 | 23.73 | 19.85 ± 0.33 | |
| 128 | 26.62 ± 0.16 | 27.80 | 24.91 ± 0.35 | |

***Table 3:*** *A comparison between radius of gyration of PETIM with that of amine functionalized PAMAM dendrimer[a] as the function of generation.*

| No of Terminal Groups | Rg (PETIM) Å | Rg (PAMAM) Å (Maiti at. al.)[19] | Rg (PAMAM) Å (Han et. al.)[35] |
|---|---|---|---|
| 4 | 5.65 | | |
| 8 | 8.90 | | 6.0 |
| 16 | 14.14 | | 8.0 |
| 32 | 17.73 | | 10.0 |
| 64 | 21.13 | 16.78 | 13.0 |
| 128 | 26.63 | 20.67 | 17.0 |
| 256 | | 26.76 | 21.3 |
| 512 | | | 27.0 |

[a] Our simulation has been done on the neutral dendrimer molecule. So we have taken the $R_g$ of PAMAM observed by Maiti et. al. [19] at high pH for the comparison. We also include the data observed by Han et. al. [35] at slightly basic pH.

***Table 4:*** *Tables contain the information about the geometric properties (Three principle moments of inertia, aspect ratio and asphericity) of PETIM dendrimer in comparison. The quantities have been averaged over 500 ps after the equilibration.*

| Generation | $<I_z>$ | $<I_y>$ | $<I_x>$ | $<I_z/I_y>$ | $<I_z/I_x>$ | $\delta$(asphericity) |
|---|---|---|---|---|---|---|
| 1(4) | 25.63 | 4.83 | 1.61 | 5.54 ± 1.34 | 20.16 ± 10.4 | 0.49 ± 0.06 |
| 2(8) | 45.30 | 25.67 | 8.43 | 1.85 ± 0.53 | 5.86 ± 2.3 | 0.17 ± 0.06 |
| 3(16) | 101.44 | 60.25 | 38.53 | 1.72 ± 0.36 | 2.70 ± 0.63 | 0.08 ± 0.04 |
| 4(32) | 173.99 | 83.13 | 57.56 | 2.19 ± 0.6 | 3.10 ± 0.75 | 0.12 ± 0.06 |
| 5(64) | 223.82 | 134.42 | 88.42 | 1.67 ± 0.26 | 2.53 ± 0.26 | 0.07 ± 0.02 |
| 6(128) | 327.20 | 232.95 | 148.87 | 1.41 ± 0.17 | 2.20 ± 0.15 | 0.04 ± 0.01 |

*Table 5: Average number of water inside dendrimer for different generation PETIM dendrimer. For criteria used in identifying the bound waters see the text. For comparison we have also shown the number water inside PAMAM dendrimer with same number of terminal group.*

| No of Terminal Groups | Bound Water (PETIM) | Bound Water (PAMAM) |
|---|---|---|
| 32 | 161 | |
| 64 | 491 | 138 |
| 128 | 1275 | 378 |
| 256 | | 890 |

**Figure Captions:**

**Figure 1**: *Two-dimensional schematic representation of fourth Generation (32 acid groups) PETIM dendrimer.*

**Figure 2:** *Variation of the Radius of gyration with time (ps) for generation 1-6 (upwards).*

**Figure 3**: *Log-linear plots of the I(q) vs. $q^2$ for (a) Petim Generation 2 (b) Petim Generation 3 (c) Petim Generation 4. The thin black lines are from the experiment and the dashed lines from MD simulation. The thick black lines are Guinier fits to the experimental data.*

**Figure 4**: *Theoretical Guinier plot (**ln(I(q)) vs $q^2$**) for different generation of PETIM dendrimer. The black lines are the linier least square fitting for the lower **q** value.*

**Figure 5:** *Plot of root mean square radius of gyration (<**Rg**>) vs generation. All the averaging has been taken over 500 ps after the equilibration. **$R_g(a)$** = radius of gyration calculated from the equation **4**. **$R_g(s)$** = radius of gyration obtained from the theoretical Guinier plot (figure 4). **$R_g(SAXS)$** = radius of gyration obtained from the experimental Guinier plot (Figure 3).*

**Figure 6**: *Radius of gyration (<**$R_g$**>) as the function of the number of monomers (atom) (in Log-Log scale), (a) for G1-G3 and (b) for generation G3-G6.*

**Figure 7:** (a) *Aspect ratio as the function of generation, **$I_z/I_x$** and **$I_z/I_y$**. (b) Asphericity has been plotted as the function of generation.*

**Figure 8:** *Instantaneous snapshots of G1-G6 PTEIM dendrimers after long MD simulations at T = 300 K. All figures are to the same scale.*

**Figure 9:** *Radial Monomer densities (arbit. unit) for different generations of EDA cored PAMAM dendrimers for all generations. Each figure is for a specific generation. The numbers shown were averaged from snapshots every 0.5 ps. The origin is at the center of mass. The last plot compares the total density profiles for all generations of PAMAM dendrimer from G1 to G6.*

**Figure 1**:

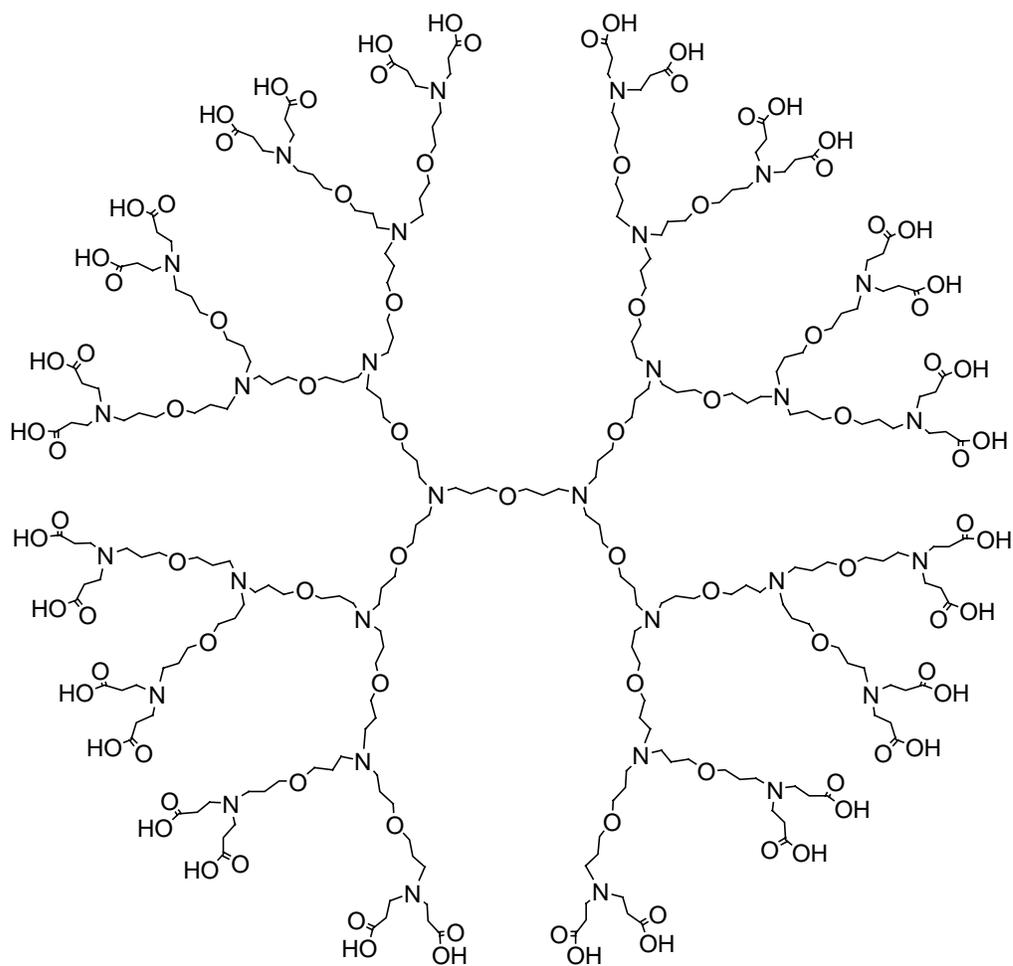

**Figure 2:**

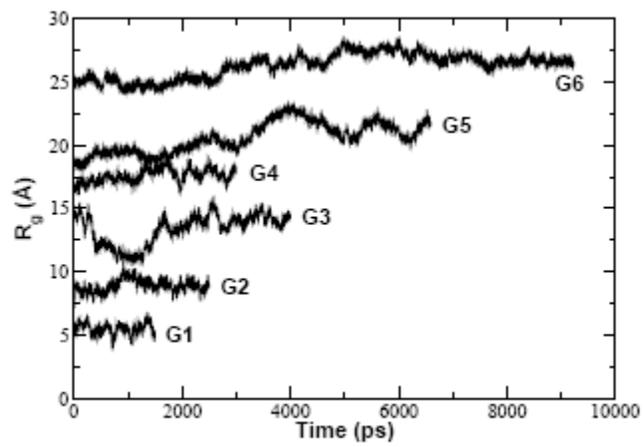

**Figure 3:**

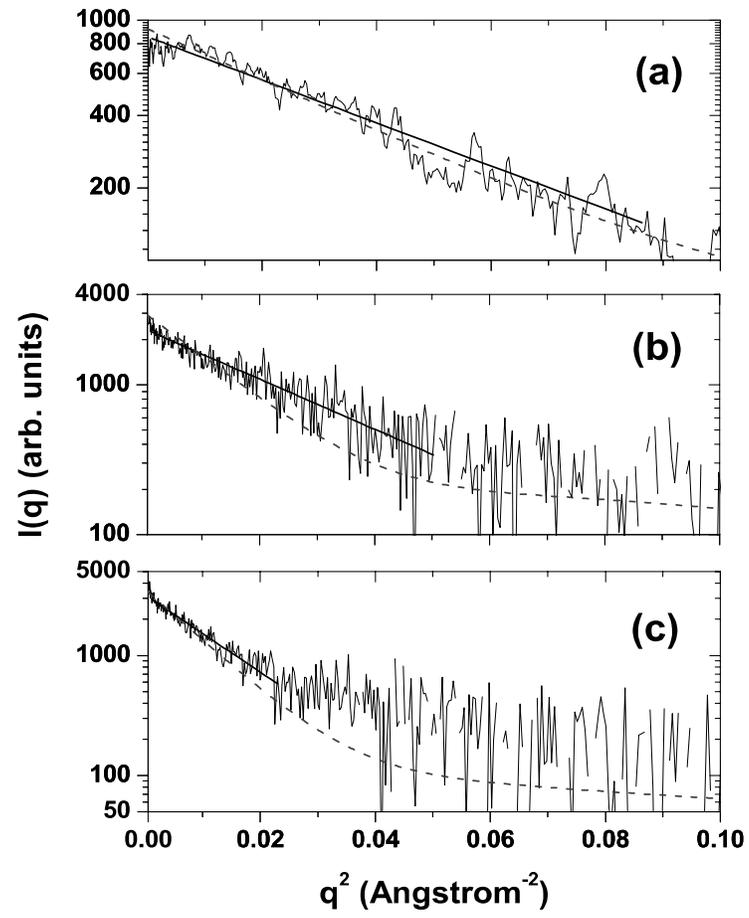

**Figure 4**:

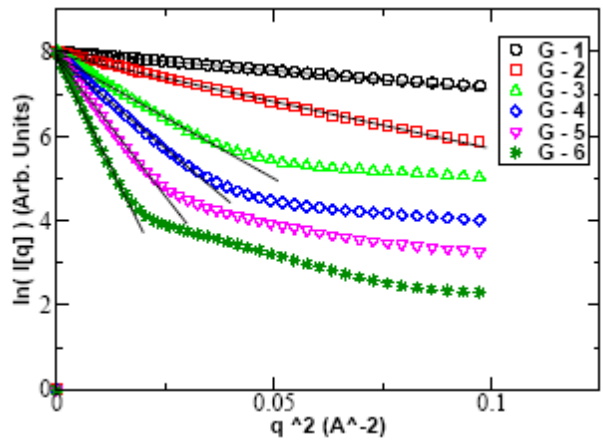

**Figure 5:**

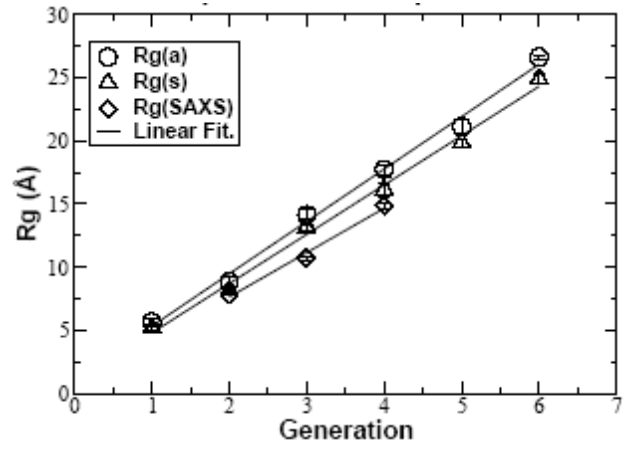

**Figure 6**:

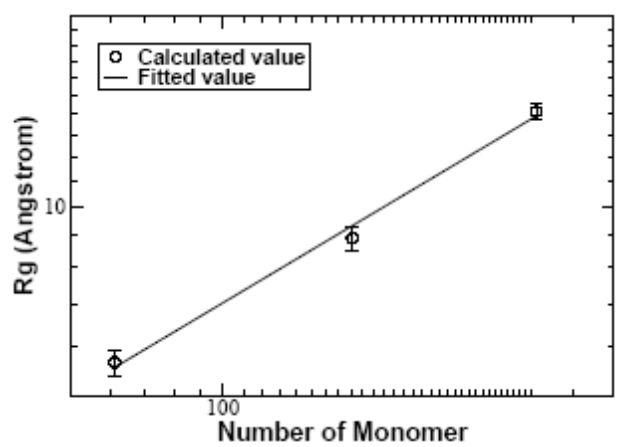 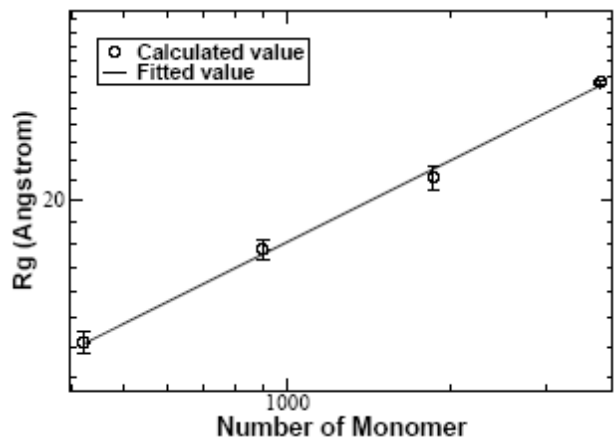

**Figure 7:**

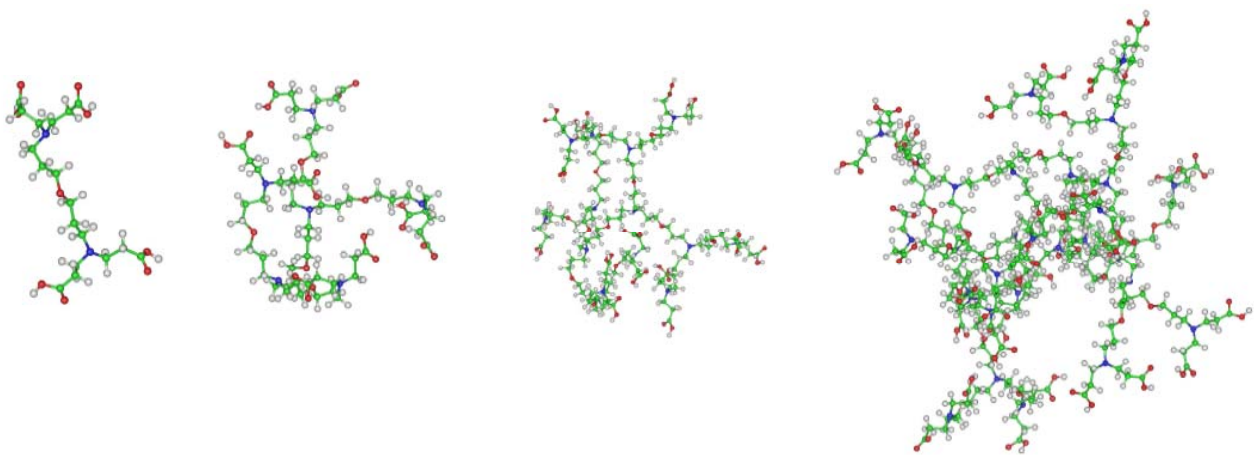

**G1**  **G2**  **G3**

**G4**

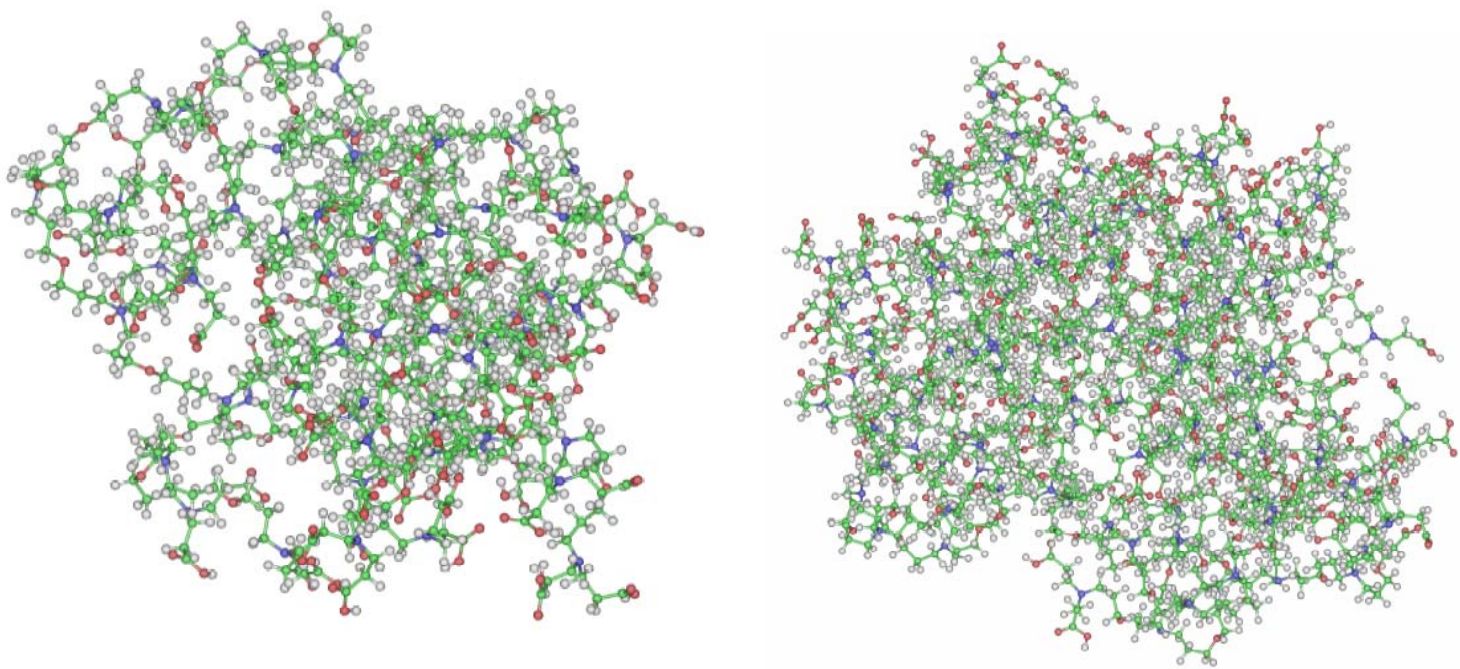

**G5**

**G6**

**Figure 8:**

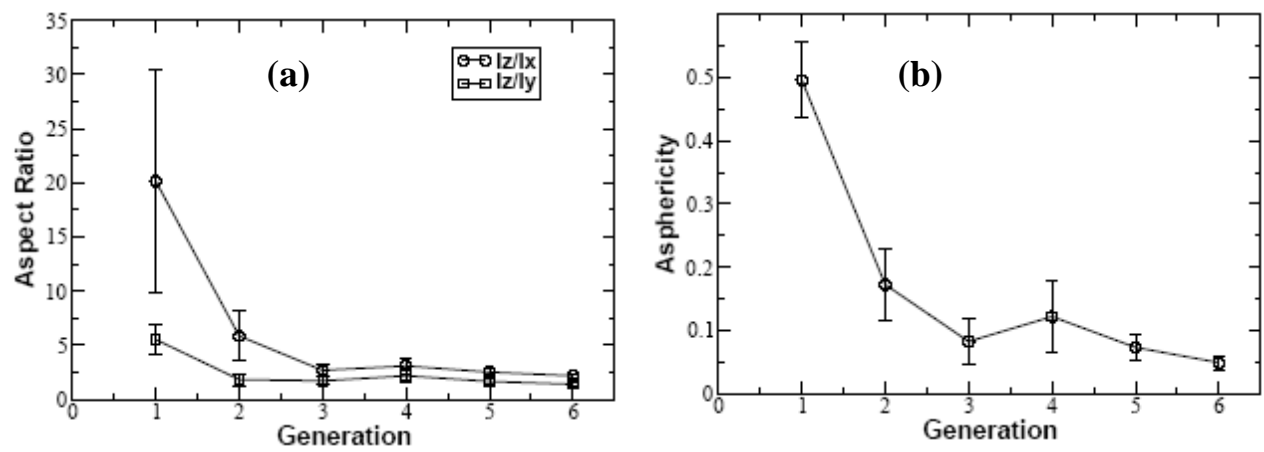

**Figure 9**

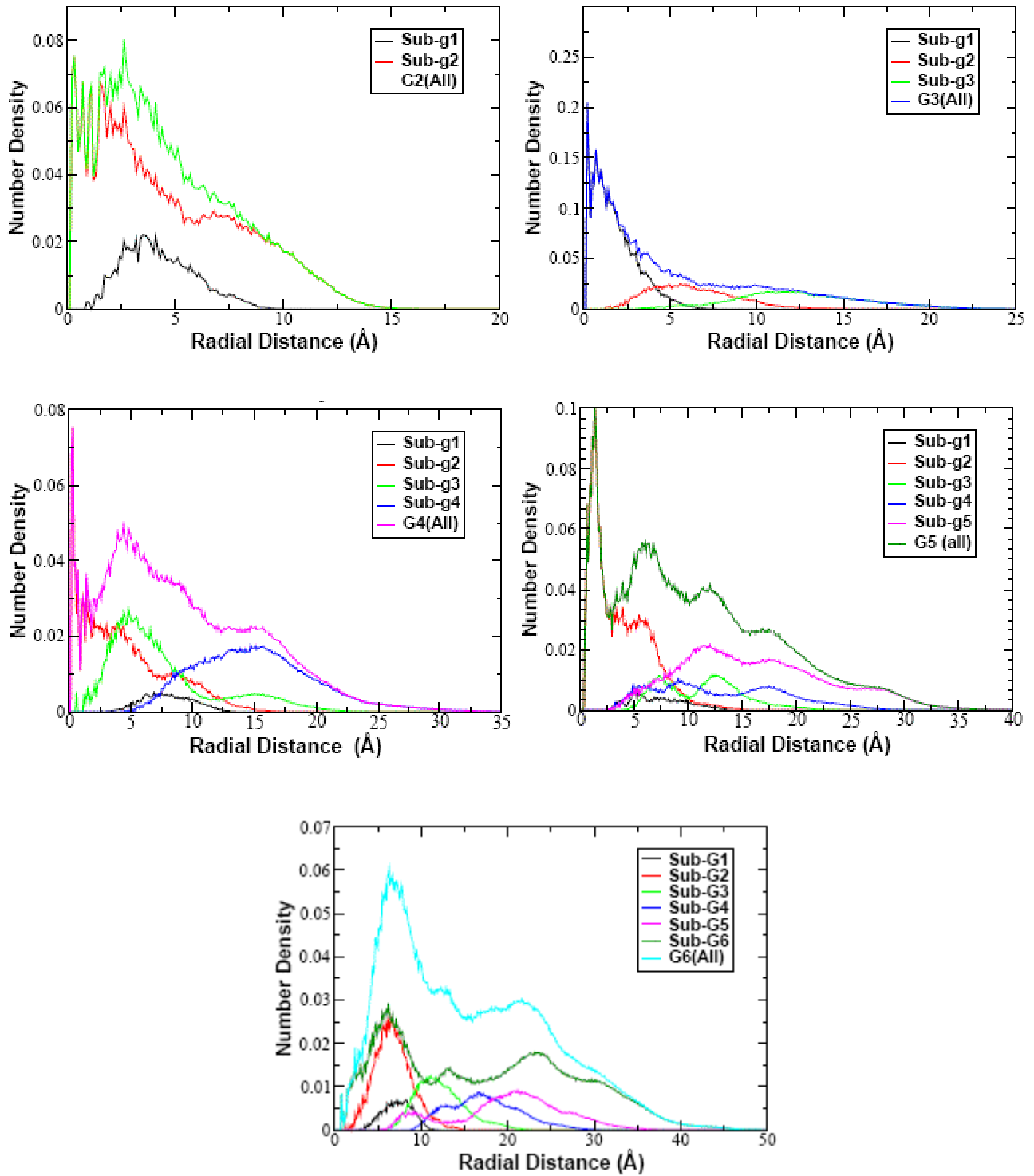